\documentclass[12pt]{article}
\usepackage{amssymb}
\usepackage{amsmath}
\usepackage{amssymb}
\usepackage{mathrsfs}
\usepackage{latexsym,bm,amsfonts}
\newtheorem{proposition}{Proposition}
\begin{document}

\title{A supersymmetric Sawada-Kotera equation}
\author{Kai Tian and Q. P. Liu\thanks{Email: qpl@cumtb.edu.cn
\; Tel: 86 10 62339015}   \\
 \em\small  Department of Mathematics, \\
\em \small China University of Mining and Technology,\\
\small\em Beijing 100083, P.R. China}

\date{}
\maketitle
\begin{abstract}
A new supersymmetric equation is proposed for the Sawada-Kotera
equation.  The integrability of this equation is shown by the
existence of Lax representation and infinite conserved quantities
and a recursion operator.
\end{abstract}

{\bf{PACS}}: 02.30.Ik; 05.45.Yv

{\bf{Key Words}}: integrability; Lax representation; recursion
operator; supersymmetry

\bigskip

\section{Introduction}
The following fifth-order evolution equation
\begin{equation}\label{sk}
u_{t}+u_{xxxxx}+5 uu_{xxx}+5 u_xu_{xx}+5 u^2u_x=0
\end{equation}
is a well-known system in soliton theory. It was proposed by Sawada
and Kotera, also by Caudrey, Dodd and Gibbon independently, more
than thirty years ago \cite{sk}\cite{cdg}, so it is referred as
Sawada-Kotera (SK) equation or Caudrey-Dodd-Gibbon-Sawada-Kotera
equation in literature. Now there are a large number of papers about
it and thus its various properties are established. For example, its
B\"{a}cklund transformation and Lax representation were given in
\cite{skaup}\cite{dodd}, its bi-Hamiltonian structure was worked out
by Fuchssteiner and Oevel \cite{fo}, and a Darboux transformation
was derived for this system \cite{aiyer}\cite{lr}, to mention just a
few (see also \cite{fg}\cite{hirota}).

Soliton equations or integrable systems have supersymmetric
analogues. Indeed, many equations such as KdV,  KP, and NLS
equations were embedded into their supersymmetric counterparts and
it turns out that these supersymmetric systems have also remarkable
properties. Thus, it is interesting to work out supersymmetric
extensions for a given integrable equation.

The aim of the Note is to propose a supersymmetric extension for
the SK equation. In this regard, we notice that Carstea
\cite{cas}, based on Hirota bilinear approach, presented the
following equation
\[
\phi_t+\phi_{xxxxx}+\left[10({\cal D}\phi)\phi_{xx}+5({\cal
D}\phi)_{xx}\phi+15({\cal D}\phi)^2\phi\right]_x=0
\]
where $\phi=\phi(x,t,\theta)$ is a fermioic super variable depending
on usual temporal variable $t$ and super spatial variables $x$ and
$\theta$. ${\cal D}=\partial_{\theta}+\theta\partial_x$ is the super
derivative. Rewriting the equation in components, it is easy to see
that this system does reduce to the SK equation when the fermionic
variable is absent. However, apart from the fact that the system can
be put into a Hirota's bilinear form, not much is known for its
integrability. We will give an alternative supersymmetric extension
for the SK equation and will show the evidence for the integrability
of our system.

The paper is organized as follows. In section two, by considering a
Lax operator and its factorization, we construct the supersymmetric
SK (sSK) equation. In section three, we will show that our sSK
equation has an interesting property, namely, it does not have the
usual bosonic conserved quantities since those, resulted from the
super residues of a fractional power for Lax operator, are trivial.
Evermore, there are infinite fermionic conserved quantities. In the
section four, we construct a recursion operator for our sSK
equation. Last section contains a brief summary of our new findings
and presents some interesting open problems.

\section{Supersymmetric Sawada-Kotera Equation}
The main purpose of this section is to construct a supersymmetric
analogy for the SK equation. To this end, we will work with the
algebra of super-pseudodifferential operators on a $(1\mid 1)$
superspace with coordinates $(x,\theta)$. We start with the
following general Lax operator
\begin{equation}\label{gen-lax}
L=\partial_x^3+\Psi\partial_x{\cal D}+U\partial_x+\Phi{\cal D}+V.
\end{equation}
By the standard fractional power method \cite{dic}, we have an
integrable hierarchy of equations given by
\begin{equation}\label{gen-sys}
\frac {\partial L}{\partial t_n}=[(L^{\frac{n}{3}})_+, L ]
\end{equation}
where we are using the standard notations:
$[A,B]=AB-(-1)^{|A||B|}BA$ is the supercommutator and the subscript
$_+$ means taking the projection to the differential part for a
given super-pseudodifferential operator. It is remarked that the
system (\ref{gen-sys}) is a kind of even order generalized SKdV
hierarchies considered in \cite{fig}.

In the following, we will consider the particular $t_5$ flow. Our
interest here is to find a minimal supersymmetric extension for
the SK equation, so we have to do reductions for the general Lax
operator (\ref{gen-lax}). To this end, we impose
\[
L+L^*=0
\]
where $^*$ means taking formal adjoint. Then  we find
\[
\Psi=0, \quad V=\frac{1}{2}(U_x-({\cal D}\Phi))
\]
that is
\[
L=\partial_x^3+U\partial_x +\Phi{\cal D}+\frac{1}{2}(U_x-({\cal
D}\Phi))
\]
a Lax operator with two field variables. In this case, we take
$B=9(L^{\frac{5}{3}})_+$, namely
\begin{eqnarray*}
B&=&9\partial_x^5+15U\partial_x^3+15\Phi{\cal
D}\partial_x^2+15(U_x+V)\partial_x^2\\
&&+15\Phi_x{\cal
D}\partial_x+(10U_{xx}+15V_x+5U^2)\partial_x\\
&&+10(\Phi_{xx}+\Phi U){\cal D}+10V_{xx}+10UV+5\Phi({\cal D}U)
\end{eqnarray*}
for convenience. Then, the flow of equations, resulted from
\[
\frac{\partial L}{\partial t}=[B,L]
\]
reads as
\begin{subequations}
\begin{eqnarray}\label{kk-sk1}
U_t+U_{xxxxx}+5\left(UU_{xx}+\frac{3}{4}U_{x}^2+\frac{1}{3}U^3+\Phi_x({\cal
D}U)+\frac{1}{2}\Phi ({\cal
D}U_x)+\frac{1}{2}\Phi\Phi_x-\frac{3}{4}({\cal D}\Phi)^2
\right)_x=0&&\\
\Phi_t+\Phi_{xxxxx}+5\left(U\Phi_{xx}+\frac{1}{2}U_{xx}\Phi+\frac{1}{2}U_x\Phi_x+U^2\Phi+
\frac{1}{2}\Phi({\cal D}\Phi_x)-\frac{1}{2}({\cal
D}\Phi)\Phi_x\right)_x=0&& \label{kk-sk2}
\end{eqnarray}
\end{subequations}
where we identify $t_5$ with $t$ for simplicity. \bigskip

\noindent {\bf Remarks}:
\begin{enumerate}
\item It is interesting
to note that the above system has an obvious reduction. Indeed,
setting $\Phi=0$, we will have the standard Kaup-Kupershimdt (KK)
equation. Therefore, we may consider it as a supersymmetric
extension of the KK equation.
\item The coupled system (\ref{kk-sk1}-\ref{kk-sk2}) admits the following simple
Hamiltonian structure
\begin{equation*}
\left(\begin{array}{c}U_t\\ \Phi_t\end{array}\right)=
\left(\begin{array}{cc} 0&\partial_x\\ \partial_x &
0\end{array}\right)\delta\mathscr{H}
\end{equation*}
where the Hamiltonian is given by
\begin{eqnarray*}
\mathscr{H} & = & \int\left[\frac{5}{4}\Phi(\mathcal{D}\phi)^2
-(\mathcal{D}U_x)(\mathcal{D}\Phi_{xx}) -\frac{5}{3}\Phi
U^3-\frac{5}{4}(\mathcal{D}U_x)(\mathcal{D}U)\Phi\right.
\\
&&~~~\left.+\frac{5}{4}(\mathcal{D}U)U_x(\mathcal{D}\Phi)+5(\mathcal{D}U)U(\mathcal{D}\Phi_x)
+\frac{5}{2}(\mathcal{D}U)\Phi_x\Phi\right]\mathrm{d}x\mathrm{d}\theta.
\end{eqnarray*}

\end{enumerate}

\bigskip

At this point, it is not clear how this system
(\ref{kk-sk1}-\ref{kk-sk2}) is related to the SK equation. To find a
supersymmetric SK equation from it, we now consider the
 factorization of the  Lax operator in the following way
\begin{eqnarray}
L&=&\partial_x^3+U\partial_x+\Phi\mathcal{D}+\frac{1}{2}(U_x-(\mathcal{D}\Phi))\nonumber\\
&=&(\mathcal{D}^3+W\mathcal{D}+\Upsilon)(\mathcal{D}^3-\mathcal{D}W+\Upsilon),\label{lax-sk}
\end{eqnarray}
which gives us a Miura-type transformation
\begin{eqnarray*}
U&=&-2W_x-W^2+(\mathcal{D}\Upsilon),\\
\Phi&=&-\Upsilon_x-2\Upsilon W,
\end{eqnarray*}
and the modified system corresponding to this factorization is given
by
\begin{equation*}
\begin{aligned}
&W_t+W_{xxxxx} + 5W_{xxx}(\mathcal{D}\Upsilon) - 5W_{xxx}W_x -
5W_{xxx}W^2 - 5W_{xx}^2 + 10W_{xx}(\mathcal{D}\Upsilon_x)\\
&~~~~ - 20W_{xx}W_xW - 5W_{xx}W(\mathcal{D}\Upsilon) - 5W_x^3 -
5W_x^2(\mathcal{D}\Upsilon) + 5W_x(\mathcal{D}\Upsilon_{xx})
+ 5W_x(\mathcal{D}\Upsilon)^2 \\
&~~~~+ 5W_xW^4 - 5W_xW(\mathcal{D}\Upsilon_x) +
10W(\mathcal{D}\Upsilon_x)(\mathcal{D}\Upsilon) -
10\Upsilon_x\Upsilon W_x + 5(\mathcal{D}W_{xx})\Upsilon_x \\
&~~~~+ 5(\mathcal{D}W_{x})\Upsilon_{xx} +
5(\mathcal{D}W_{x})\Upsilon W_x + 10(\mathcal{D}W_{x})\Upsilon W^2 -
5(\mathcal{D}W)\Upsilon_xW_x+ 10(\mathcal{D}W)\Upsilon_xW^2\\
&~~~~ + 10(\mathcal{D}W)\Upsilon (\mathcal{D}\Upsilon_x) -
5(\mathcal{D}W)\Upsilon W_{xx} +
30(\mathcal{D}W)\Upsilon W_xW=0,\\[10pt]
&\Upsilon_t+\Upsilon_{xxxxx} + 5\Upsilon_{xxx}(\mathcal{D}\Upsilon)
- 5\Upsilon_{xxx}W^2 + 5\Upsilon_{xx}(\mathcal{D}\Upsilon_x) +
5\Upsilon_{xx}W_{xx} - 25\Upsilon_{xx}W_xW \\
&~~~~+ 5\Upsilon_{xx}W(\mathcal{D}\Upsilon) +
5\Upsilon_{x}(\mathcal{D}\Upsilon)^2 + 5\Upsilon_{x}W_{xxx} -
25\Upsilon_{x}W_{xx}W - 25\Upsilon_{x}W_x^2 +
5\Upsilon_{x}W_x(\mathcal{D}\Upsilon)\\
&~~~~ + 10\Upsilon_{x}W_xW^2 + 5\Upsilon_{x}W^4 -
10\Upsilon_{x}W^2(\mathcal{D}\Upsilon) +
5\Upsilon_{x}W(\mathcal{D}\Upsilon_x) - 10\Upsilon W_{xxx}W -
20\Upsilon W_{xx}W_x \\
&~~~~+ 10\Upsilon W_{xx}W^2 + 30\Upsilon W_{x}^2W + 20\Upsilon
W_{x}W^3 - 30\Upsilon W_{x}W(\mathcal{D}\Upsilon) - 10\Upsilon
W^2(\mathcal{D}\Upsilon_x) \\&~~~~-
5(\mathcal{D}W_x)\Upsilon_{x}\Upsilon -
5(\mathcal{D}W)\Upsilon_{xx}\Upsilon -
10(\mathcal{D}W)\Upsilon_{x}\Upsilon W=0.
\end{aligned}
\end{equation*}

Although this modification does indeed have a complicated form, the
remarkable fact is that it allows a simple reduction. What we need
to do is simply putting $W$ to zero, namely
\[
W=0, \quad \Upsilon=\phi.
\]
In this case, we have
\begin{equation}\label{ssk} \phi_{t} +\phi_{xxxxx}+5\phi_{xxx}(D\phi)+
      5\phi_{xx}(D\phi_x)+5\phi_x(D\phi)^2=0
\end{equation}
this equation is our supersymmetric SK equation.  To see the
connection with the original SK equation (\ref{sk}), we let
$\phi=\theta u(x,t)+\xi(x,t)$ and write the equation (\ref{ssk}) out
in components
\begin{subequations}
\begin{eqnarray}
u_t  + u_{xxxxx}+5 uu_{xxx}+5 u_xu_{xx}
      +5u^2u_x- 5\xi_{xxx}\xi_x&=&0,\\
\xi_t+\xi_{xxxxx}+5 u\xi_{xxx}+ 5 u_x\xi_{xx}+
      5u^2\xi_{x}&=&0.
\end{eqnarray}
\end{subequations}
It is now obvious that the system reduces to the SK equation when
$\xi=0$. Therefore, our system (\ref{ssk}) does qualify as a
supersymmetric SK equation.

Our system (\ref{ssk}) is integrable in the sense that it has a Lax
representation. In fact, the factorization (\ref{lax-sk}) implies
that the reduced Lax operator has the following appealing form
\begin{equation}\label{sklax}
L=(\mathcal{D}^3+\phi)(\mathcal{D}^3+\phi)
\end{equation}
or
\begin{equation*}
L=\partial_x^3+(\mathcal{D}\phi)\partial_x-\phi_x\mathcal{D}
+(\mathcal{D}\phi_x)
\end{equation*}

\section{Conserved Quantities}
In general, an integrable system has infinite number of conserved
quantities. Since the sSK equation has a simple Lax operator
(\ref{sklax}), it is natural to take advantage of the fractional
power method of Gel'fand and Dickey  \cite{dic}  to find conserved
quantities. In the present situation, we have to work with the super
residue of a pesudodifferential operator.

The obvious choice in this case is to consider the operators
$L^{\frac{n}{3}}$ and their super residues.  Then, we have the

\begin{proposition}
\begin{equation*}
\mathrm{sres}L^{\frac{n}{3}}\in\mathrm{Im}\mathcal{D}~.
\end{equation*}
where $\mathrm{sres}$ means taking the super residue of a super
pseudodifferential operator.
\end{proposition}

\noindent
Proof: As observed already in \cite{mr}, there exists a
unique odd operator $\Lambda=\mathcal{D}+\mathrm{O}(1)$, whose
coefficients are all differential polynomials of $\phi$, such that
\begin{equation*}
(\mathcal{D}^3+\phi)=\Lambda^3,
\end{equation*}
thus, the Lax operator (\ref{sklax}) is written as
\begin{equation*}
L=(\mathcal{D}^3+\phi)(\mathcal{D}^3+\phi)=\Lambda^6.
\end{equation*}
From it we have
\begin{equation*}
\mathrm{sres}L^{\frac{n}{3}}=\mathrm{sres}\Lambda^{2n}
=\frac{1}{2}\mathrm{sres}\{\Lambda^{2n-1}\Lambda +\Lambda
\Lambda^{2n-1}\} =\frac{1}{2}\mathrm{sres}[\Lambda^{2k-1},\Lambda
]\in \mathrm{Im}\mathcal{D}~.
\end{equation*}
This completes the proof.
\bigskip

\noindent
\textbf{Remark:} The triviality of $L^{\frac{n}{3}}$
implies that the Lax operator could not generate any Hamiltonian
structures for the equation (\ref{ssk}).
\bigskip

To find nontrivial conserved quantities, we now turn to
$L^{\frac{n}{6}}$ rather than $L^{\frac{n}{3}}$.  It is easy to
prove  that
\begin{equation*}
\frac{\partial}{\partial
t}L^{\frac{1}{6}}=[9(L^{\frac{5}{3}})_+,L^{\frac{1}{6}}]
\end{equation*}
thus
\begin{equation*}
\frac{\partial}{\partial
t}L^{\frac{n}{6}}=[9(L^{\frac{5}{3}})_+,L^{\frac{n}{6}}].
\end{equation*}
Consequently, the super residue of $L^{\frac{n}{6}}$ is conserved.

By direct calculation, we obtain the first two nontrivial conserved
quantities
\begin{eqnarray*}
\int\mathrm{sres}L^{\frac{7}{6}}\mathrm{d}x\mathrm{d}\theta
&=&-\frac{1}{9}\int[2(\mathcal{D}\phi_{xx})+(\mathcal{D}\phi)^2
-6\phi_x\phi]\mathrm{d}x\mathrm{d}\theta\\
\int\mathrm{sres}L^{\frac{11}{6}}\mathrm{d}x\mathrm{d}\theta
&=&-\frac{1}{81}\int[6(\mathcal{D}\phi_{xxxx})+18(\mathcal{D}\phi_{xx})(\mathcal{D}\phi)
+9(\mathcal{D}\phi_x)^2\\
&&~~~~~~~~~+4(\mathcal{D}\phi)^3-18\phi_{xxx}\phi+6\phi_{xx}\phi_x
-36\phi_x\phi(\mathcal{D}\phi)]\mathrm{d}x\mathrm{d}\theta\\
\end{eqnarray*}

\noindent {\textbf{Remarks:}}
\begin{enumerate}
\item
What is remarkable is that the conserved quantities found in this
way, unlike the supersymmetric KdV case \cite{lm}\cite{kers}, are
local.
\item All those conserved quantities are fermioic. To our knowledge,
this is the first supersymmetric integrable system whose only
conserved quantities are fermioic.
\end{enumerate}

\section{Recursion operator}
An integrable system often appears as a particular flow of hierarchy
equations and an important ingredient in this aspect is the
existence of recursion operators. In this section, we deduce the
recursion operator for the sSK equation (\ref{ssk}) following  the
method proposed   in \cite{sok}. We first notice that the sSK
hierarchy can be written as
\begin{equation}\label{sklax2}
\frac{\partial}{\partial t_n}L=[(L^{\frac{n}{3}})_+,L]
\end{equation}
where $L$ is given by (\ref{sklax}). It is easy to see that the flow
equations are nontrivial only if $n$ is an integer satisfying
\begin{equation*}
n\neq 0~\mathrm{mod}~3\quad\text{and}\quad n=1~\mathrm{mod}~2.
\end{equation*}
Therefore the next flow which is achieved by applying recursion
operator to (\ref{sklax2}) should be
\begin{equation}
\frac{\partial}{\partial t_{n+6}}L=[(L^{\frac{n+6}{3}})_+,L].
\end{equation}
But
\begin{equation*}
\begin{aligned}
\left[(L^{\frac{n+6}{3}})_+,L\right]
&=\left[\left(L^2(L^{\frac{n}{3}})_+ +
L^2(L^{\frac{n}{3}})_-\right)_+,L\right]\\
&=\left[L^2(L^{\frac{n}{3}})_+,L\right]+\left[\left(L^2(L^{\frac{n}{3}})_-\right)_+,L\right]\\
&=L^2\left[(L^{\frac{n}{3}})_+,L\right]+[R_n,L]\\
&=L^2\frac{\partial}{\partial t_{n}}L+[R_n,L]
\end{aligned}
\end{equation*}
where
\begin{equation}\label{rec-fm}
R_n=\left(L^2(L^{\frac{n}{3}})_-\right)_+
\end{equation}
is a differential operator of $O(\partial_x^5\mathcal{D})$, that is,
\begin{eqnarray*}
R_n&=(\alpha\partial_x^5+\beta\partial_x^4+\gamma\partial_x^3+\delta\partial_x^2+\xi\partial_x
+\eta)\mathcal{D}\\
&+a\partial_x^5+b\partial_x^4+c\partial_x^3+d\partial_x^2+e\partial_x+f.
\end{eqnarray*}

Therefore,
\begin{equation}\label{re-flw}
\frac{\partial}{\partial t_{n+6}}L=L^2\frac{\partial}{\partial
t_{n}}L+[R_n,L].
\end{equation}

Next we may determine the coefficients in $R_n$. Using
(\ref{re-flw}), we obtain
\begin{equation*}
\begin{aligned}
&a=\frac{1}{3}\underline{(\mathcal{D}^{-1}\phi_n)},\qquad
b=2(\mathcal{D}\phi_n), \\
&c=\frac{44}{9}(\mathcal{D}\phi_{n,x})+\frac{5}{3}(\mathcal{D}\phi)(\mathcal{D}^{-1}\phi_n)
+\frac{4}{9}\underline{(\partial_x^{-1}\phi_x\phi_n)}, \\
&d=\frac{55}{9}(\mathcal{D}\phi_{n,xx})+\frac{19}{9}(\mathcal{D}\phi)(\mathcal{D}\phi_n)
+\frac{5}{9}\phi_x\phi_n+\frac{10}{9}(\mathcal{D}\phi_x)(\mathcal{D}^{-1}\phi_n) \\
&e=\frac{1}{27}\{106(\mathcal{D}\phi_{n,xxx})+74(\mathcal{D}\phi)(\mathcal{D}\phi_{n,x}),
-14\phi_x\phi_{n,x}+79(\mathcal{D}\phi_x)(\mathcal{D}\phi_n) \\
&\quad\;\;+27\phi_{xx}\phi_n+[23(\mathcal{D}\phi_{xx})
+4(\mathcal{D}\phi)^2](\mathcal{D}^{-1}\phi_n)+16(\mathcal{D}\phi)(\partial_x^{-1}\phi_x\phi_n) \\
&\quad\;\;+2\mathcal{D}^{-1}[(\phi_{xxx}+\phi_x(\mathcal{D}\phi))(\mathcal{D}^{-1}\phi_n)
-3(\mathcal{D}\phi)(\mathcal{D}^{-1}\phi_x\phi_n) \\
&~~~~~~~~~~~~~-2\phi_x(\partial_x^{-1}\phi_x\phi_n)+2\mathcal{D}^{-1}(\phi_{xxx}\phi_n
+2\phi_x(\mathcal{D}\phi)\phi_n)]\}, \\
&f=\frac{1}{27}\{28(\mathcal{D}\phi_{n,xxxx})+32(\mathcal{D}\phi)(\mathcal{D}\phi_{n,xx})-20\phi_x\phi_{n,xx}
+54(\mathcal{D}\phi_x)(\mathcal{D}\phi_{n,x}) \\
&\quad\;\;+16\phi_{xx}\phi_{n,x}+[30(\mathcal{D}\phi_{xx})
+4(\mathcal{D}\phi)^2](\mathcal{D}\phi_n)+[8\phi_{xxx}+4\phi_x(\mathcal{D}\phi)]\phi_n \\
&\quad\;\;+[10(\mathcal{D}\phi_{xxx})+10(\mathcal{D}\phi_x)(\mathcal{D}\phi)](\mathcal{D}^{-1}\phi_n)
-8\phi_x\underline{(\mathcal{D}^{-1}\phi_x\phi_n)} \\
&\quad\;\;+12(\mathcal{D}\phi_x)(\partial_x^{-1}\phi_x\phi_n)\},
\end{aligned}
\end{equation*}
\begin{equation*}
\begin{aligned}
&\alpha=0, \;\;\beta=-\frac{1}{3}\phi_n,\;\;\gamma=\frac{5}{3}\phi_{n,x},\\
&\delta=-\frac{1}{9}\{29\phi_{n,xx}+5\phi_n(\mathcal{D}\phi)+5\phi_x(\mathcal{D}^{-1}\phi_n)
-2(\mathcal{D}^{-1}\phi_x\phi_n)\},\\
&\xi=-\frac{1}{9}\{26\phi_{n,xxx}+16\phi_x(\mathcal{D}\phi_n)+3\phi_n(\mathcal{D}\phi_x)
+14\phi_{n,x}(\mathcal{D}\phi)+5\phi_{xx}(\mathcal{D}^{-1}\phi_n)\},\\
&\eta=-\frac{1}{27}\{28\phi_{n,xxxx}+32(\mathcal{D}\phi)\phi_{n,xx}+28\phi_x(\mathcal{D}\phi_{n,x})
+26(\mathcal{D}\phi_x)\phi_{n,x}\\
&\quad\;\;+28\phi_{xx}(\mathcal{D}\phi_n)+[2(\mathcal{D}\phi_{xx})
+4(\mathcal{D}\phi)^2]\phi_n+[10\phi_{xxx}+10\phi_x(\mathcal{D}\phi)](\mathcal{D}^{-1}\phi_n)\\
&\quad\;\;-2(\mathcal{D}\phi)(\mathcal{D}^{-1}\phi_x\phi_n)+12\phi_x(\partial_x^{-1}\phi_x\phi_n)
-2\underline{\mathcal{D}^{-1}[\phi_{xxx}\phi_n+2\phi_x(\mathcal{D}\phi)\phi_n]}\}.
\end{aligned}
\end{equation*}
where we used the shorthand notation $\phi_n=\partial\phi/\partial
t_n$.

Finally, we have the recursion operator
\begin{equation*}
\begin{aligned}
\mathscr{R}=&\partial_x^6+6(\mathcal{D}\phi)\partial_x^4+9(\mathcal{D}\phi_x)\partial_x^3
+6\phi_{xx}\partial_x^2\mathcal{D}
+\{5(\mathcal{D}\phi_{xx})+9(\mathcal{D}\phi)^2\}\partial_x^2\\
&+\{9\phi_{xxx}+12\phi_x(\mathcal{D}\phi)\}\partial_x\mathcal{D}
+\{(\mathcal{D}\phi_{xxx})+9(\mathcal{D}\phi_x)(\mathcal{D}\Phi)\}\partial_x\\
&+\{5\phi_{xxxx}+12\phi_{xx}(\mathcal{D}\phi)+6\phi_x(\mathcal{D}\phi_x)\}\mathcal{D}
+\{4(\mathcal{D}\phi_{xx})(\mathcal{D}\phi)+4(\mathcal{D}\phi)^3-3\phi_{xx}\phi_x\}\\
&+\{\phi_{xxxxx}+5\phi_{xxx}(\mathcal{D}\phi)+5\phi_{xx}(\mathcal{D}\phi_x)
+2\phi_x(\mathcal{D}\phi_{xx})+6\phi_x(\mathcal{D}\phi)^2\}{\mathcal{D}^{-1}}\\
&-\{2(\mathcal{D}\phi_{xx})+2(\mathcal{D}\phi)^2\}{\mathcal{D}^{-1}\phi_x}
-4\phi_x(\mathcal{D}\phi){\partial_x^{-1}\phi_x}
-2(\mathcal{D}\phi){\mathcal{D}^{-1}[\phi_{xxx}+2\phi_x(\mathcal{D}\phi)]}\\
&-2\phi_x{\mathcal{D}^{-1}\{(\phi_{xxx}+\phi_x(\mathcal{D}\phi))\mathcal{D}^{-1}
-3(\mathcal{D}\phi)\mathcal{D}^{-1}\phi_x
-2\phi_x\partial_x^{-1}\phi_x+2\mathcal{D}^{-1}[\phi_{xxx}+2\phi_x(\mathcal{D}\phi)]\}}
\end{aligned}
\end{equation*}

\noindent \textbf{Remark:} When calculating the coefficients of
$R_n$, one should solve a system of differential equations. Due to
nonlocality (those underlined terms), there is certain ambiguity and
to avoid it, we used the $t_7$-flow
\begin{eqnarray*}
\phi_{t_7}&=&\phi_{xxxxxxx}+7\phi_{xxxxx}(\mathcal{D}\phi)
+14\phi_{xxxx}(\mathcal{D}\phi_x)+14\phi_{xxx}(\mathcal{D}\phi_{xx})\\
&&+14\phi_{xxx}(\mathcal{D}\phi)^2+7\phi_{xx}(\mathcal{D}\phi_{xxx})
+28\phi_{xx}(\mathcal{D}\phi_x)(\mathcal{D}\phi)\\
&&+14\phi_x(\mathcal{D}\phi_{xx})(\mathcal{D}\phi)
+7\phi_x(\mathcal{D}\phi_x)^2+\frac{28}{3}\phi_x(\mathcal{D}\phi)^3~.
\end{eqnarray*}

\section{Conclusion}
Summarizing, we find a supersymmetric SK equation which has Lax
representation. We also obtain infinite conserved quantities and a
recursion operator for this new proposed system. These imply that
the system is integrable. It is interesting to establish other
properties for it, such as B\a"{a}cklund transformation, Hirota
bilinear form, etc..
\bigskip

\noindent
{\textbf{ Acknowledgements}} The calculations were done
with the assistance of SUSY2 package of Popowicz \cite{pop}.
 We would like
 to thank him for helpful discussion about his package. The comments of anonymous referee has been very useful.
 The work is supported in part by National
Natural Science Foundation of China under the grant numbers
10671206 and 10731080.

\end{document}